\documentclass[11pt,a4paper,notitlepage,superscriptaddress,nofootinbib]{article}

\usepackage{jcappub}
\usepackage{lmodern}
\usepackage{verbatim}

\usepackage[T1]{fontenc}
\usepackage[utf8]{inputenc}
\usepackage{color}
\usepackage{subcaption}
\usepackage{amsmath}
\usepackage{amssymb}
\usepackage{graphicx}
\usepackage{tabularx}
\usepackage{esint}

\begin{document}

\title{Transient weak gravity in scalar-tensor theories}

\author[1]{Manuel Wittner,}
\author[1]{Giorgio Laverda,}
\author[1,2]{Oliver F. Piattella}
\author[1]{and Luca Amendola}

\affiliation[1]{Institut f{\"u}r Theoretische Physik, Ruprecht-Karls-Universit{\"a}t Heidelberg,  Philosophenweg 16, 69120 Heidelberg, Germany}
\affiliation[2]{Department of Physics, Universidade Federal do Esp\'irito Santo,
Avenida Fernando Ferrari 514, 29075-910 Vit\'oria, Esp\'irito Santo, Brazil}

\emailAdd{wittner@thphys.uni-heidelberg.de}
\emailAdd{laverda@thphys.uni-heidelberg.de}
\emailAdd{oliver.piattella@cosmo-ufes.org}
\emailAdd{l.amendola@thphys.uni-heidelberg.de}

\abstract{
Sub-horizon perturbations in scalar-tensor theories have been shown to grow generically faster than in uncoupled models due to a positive, additive Yukawa force. 
In such cases, the amount of clustering becomes larger than in the standard cosmological model, exacerbating the observed tension in the $\sigma_8$ parameter.
Here we show instead that in some simple cases of conformally coupled dark energy one can obtain a transient regime of negative Yukawa force, without introducing ghosts or other instabilities.}

\keywords{Horndeski theories, coupled quintessence, modified gravity}

\maketitle


\section{Introduction}

The search for alternative theories of gravity describing new observable phenomena in cosmology has been in the focus of a large amount of work in the last decade. This field of research is partially driven by theoretical puzzles related to the cosmological constant, such as the problem of coincidence \cite{Zlatev:1998tr, Velten:2014nra} and fine-tuning \cite{Weinberg:1987dv, Weinberg:1988cp} and partially by the availability of new large data sets that allow to push our tests of General Relativity from the Solar System to the ``realm of the nebulae''. More recently, puzzling observations like the $H_0$ discrepancy \cite{Ade:2015xua,Riess_2018} and the $\sigma_8$ tension \cite{Heymans_2012,Hildebrandt_2016} have also motivated the exploration of alternatives to standard gravity. 

One particularly popular set of alternatives to Einstein gravity is represented by scalar-tensor theories of gravity, in which the general-relativistic, tensorial nature of gravity is extended to incorporate also a scalar field $\phi$. The most studied forms of such theories are embodied in the Horndeski theory and its extensions \cite{Horndeski:1974wa, Bellini:2014fua, Gleyzes:2014dya, Langlois:2015cwa}. Horndeski models comprise the most general equations of motion that contain at most second-order derivatives of the fields and their extensions such as DHOST (degenerate higher-order scalar-tensor theories) have equations of motions with higher-than-second-order derivatives but nonetheless are free from Ostrogradski instabilities \cite{Ostrogradsky:1850fid, Woodard:2006nt, Trodden:2016zcu, Langlois:2018dxi}. 
Among the new observable phenomena introduced by these theories we can mention the time-dependence of the speed of gravitational waves, the non-vanishing anisotropic stress, the growth of fluctuations in the radiation era \cite{Amendola:2017xhl}, the emergence of early dark energy, a screening mechanism in high-density regions \cite{Khoury:2003rn}, the scale-dependent growth rate of small dark matter fluctuations. All these features are in principle testable and constitute one of the main aims of future surveys such as Euclid.\footnote{See e.g. \url{https://sci.esa.int/web/euclid}.}

In this paper we investigate in some detail the growth of matter fluctuations in a simple scalar-tensor theory characterised by a conformal coupling. 
Even in the simplest case of growth in the dark matter-dominated epoch, the problem can only be treated numerically. To this purpose, publicly available codes such as Hi-Class and EFTCamb \cite{Zumalacarregui:2016pph, Hu:2013twa} are a precious tool in order to obtain theoretical predictions. On the other hand, the most relevant regime for large-scale galaxy observations is the late-times Universe at sub-horizon scales. In this case, a good approximation for the evolution equations at small scales, at least for that class of models for which the scalar-field sound speed is not too small, is the so-called quasi-static approximation \cite{2011PhLB..706..123D}. In this approximation one assumes that $k/aH \to \infty$ and keeps in the evolution equations only the dominant terms in this quantity. It turns out that small dark matter fluctuations then satisfy an evolution equation similar to the standard one obtained in General Relativity, but with an effective gravitational coupling $G_{\rm eff}=G_\text{N}[1+Y(k)]$, where  $G_\text{N}$ is the usual Newton gravitational constant  in absence of the scalar coupling, and where the Yukawa correction $Y$ has the following form:
\begin{equation}
   Y = 2M_\text{P}^2 Q^2\frac{k^2}{k^2 + M^2}\;, \label{eq:qsa1}
\end{equation}
($M_\text{P}$ being the Planck mass) i.e. a ratio of quadratic polynomials in $k$, where $Q$ is the scalar coupling strength and $M$ the field mass times the scale factor squared. The coefficients  $Q$ and $M$ are functions of the background quantities and their explicit forms in the case of Horndeski theories can be found e.g. in Refs.~ \cite{2011PhLB..706..123D, Amendola:2019laa}.
In real space, such an effective Newton's constant gives rise to the Newton-Yukawa potential
\begin{equation}
    V(r)=-\frac{G_\text{N} m}{r}(1+2 Q^2 e^{-Mr})\;.
\end{equation}
Since  matter fluctuations obey a Poisson equation with a $G_{\rm eff}=G(k)$, they will in general acquire a scale dependence, in contrast to the standard case of General Relativity. Evidences of this scale-dependence can be in principle observed in the galaxy distribution as e.g. the scale-dependent combination $f\sigma_8(z)$ (see e.g. Ref.~\cite{Perenon:2016blf}), thereby providing important information on the scalar sector of gravity.

The detailed form of $Y(k)$ depends, of course, on the particular model. However, it is generically found for $k \to \infty$ that $Y > 0$ \cite{Amendola:2017orw}, unless the field equation of state $w_{\phi}$ is in the phantom regime, $w_{\phi}<-1$. This can be achieved, for instance, with a non-standard kinetic energy, which however introduces instabilities in the model \cite{Amendola:2004qb}. 

In this paper we analyse one of the simplest forms of scalar-tensor theories, namely a dark matter sector minimally coupled to a conformally transformed metric, with a conformal factor $C(\phi)$. We also take a standard kinetic  term for the field and a simple quadratic field potential, so we do not introduce any  instability.   We find that, under certain conditions, the second derivative of $C$ with respect to $\phi$ acts a second mass scale, beside the scalar field mass. If this second mass scale is not negligible with respect to $k^2$, it  modifies the effective gravitational coupling. Interestingly, we find that the Yukawa correction might become negative, thereby inducing a weak effective gravitational force.  This regime is in general a transient one, and the duration and strength will depend on the particular model. At the same time, we find another new feature in the quasi-static regime, namely a $k$-dependence of the friction term. We also show that the following square-exponential model,
\begin{equation}\label{csqexp}
    C(\phi) = e^{m_C^{-2}\phi^2}\;,
\end{equation}
where $m_C$ is an arbitrary mass scale,  is particularly suitable for inducing the novel behaviour of the effective gravitational coupling. 
For illustrative purposes, we therefore specialise some of our expressions to this case.

The interest of our finding is that a weaker gravity leads, in general, to a weaker fluctuation growth. This might explain why some observations find a lower clustering strength $\sigma_8$ as compared to $\Lambda$CDM \cite{MacCrann:2014wfa,Ade:2015fva,Joudaki:2016mvz,Abbott:2017wau}. However, in this paper we do not explore in any detail the cosmological data and only confine ourselves to demonstrating the existence of a transient weak gravity regime.


\section{The model}\label{Sec:Model}

The model under investigation is described by the following action, see e.g.  Ref.~\cite{vandeBruck:2015ida}: 
\begin{equation}\label{action}
	\mathcal S = \int d^4x\sqrt{-g}\left[\frac{M_\text{P}^2}{2}R + L(g,\phi) + L_{\rm SM}(g,\psi_{\rm SM})\right] + \int d^4x\sqrt{-\tilde{g}}\tilde{L}_\text{m}(\tilde g,\psi_\text{m})\;.
\end{equation}
The gravitational sector is the usual one, described by the Einstein-Hilbert action, to which a canonical scalar field $\phi$ and Standard-Model (SM) matter fields, generically referred to as $\psi_{\rm SM}$, minimally couple. The Dark Matter (DM) field (or fields), indicated as $\psi_\text{m}$,  couples instead to an effective geometry, characterised by a metric $\tilde g$, which is related to $g$ by the following disformal transformation:
\begin{equation}\label{disftransf}
	\tilde g_{\mu\nu} = C(\phi)g_{\mu\nu} + D(\phi)\phi_{,\mu}\phi_{,\nu}\;,
\end{equation}
where $C$ and $D$ are generic functions of the scalar field $\phi$. The latter is governed by the following Lagrangian:
\begin{equation}
	L = -\frac{1}{2}g^{\mu\nu}\phi_{,\mu}\phi_{,\nu} - V(\phi)\;,
\end{equation}
where $V(\phi)$ is some potential (from now on, we normally drop the explicit dependence of $C$, $D$ and $V$ on $\phi$). As it appears  from Eq.~\eqref{action}, we work in the Einstein frame and assume baryons to be decoupled from the scalar field. This choice has the advantage that we do not need to invoke any particular screening mechanism to pass the local gravity constraints and we do not have to worry about the gravitational waves speed constraints \cite{Monitor:2017mdv}. Moreover, since baryons are a relatively minor component, we assume for simplicity that their density is actually negligible with respect to coupled matter and scalar field.

The general, covariant Einstein field equations for such a model are given by \cite{Mifsud:2017fsy}
\begin{equation}
    R_{\mu\nu}-\frac{1}{2}g_{\mu\nu}R = M_\text{P}^{-2} \left( T^\phi_{\mu\nu}+T^\text{SM}_{\mu\nu} + T^\text{m}_{\mu\nu}  \right)\;,
\end{equation}
where the energy-momentum tensors of the different sectors are defined as
\begin{align}
    T^\phi_{\mu\nu} &\equiv \phi_{,\mu}\phi_{,\nu}-g_{\mu\nu}\left( \frac{1}{2} g^{\rho\sigma}\phi_{,\rho}\phi_{,\sigma}+V \right)\;, \\
    T^\text{SM}_{\mu\nu} &\equiv -\frac{2}{\sqrt{-g}} \frac{\delta(\sqrt{-g}L_\text{SM})}{\delta g^{\mu\nu}}\;,\\
    T^\text{m}_{\mu\nu} &\equiv -\frac{2}{\sqrt{-g}} \frac{\delta(\sqrt{-\tilde{g}}\tilde{L}_\text{m})}{\delta g^{\mu\nu}}\;.
\end{align}
The SM particles are minimally coupled and are therefore described by the standard conservation equation
\begin{equation}
    \nabla^\mu T^\text{SM}_{\mu\nu} = 0\;.
\end{equation}
On the other hand, due to the coupling between $\phi$ and dark matter, their conservation equation only holds for the total energy-momentum tensor \cite{Zumalacarregui:2012us}
\begin{equation}
    \nabla^\mu \left( T^\phi_{\mu\nu}+T^\text{m}_{\mu\nu} \right) = 0\;,
\end{equation}
while the individual components are related by
\begin{equation}
    -\nabla^\mu T^\phi_{\mu\nu} = \nabla^\mu T^\text{m}_{\mu\nu} \equiv \Gamma \phi_{,\nu}\;,
\end{equation}
where the coupling function is given by
\begin{equation}
    \Gamma = \frac{C_{,\phi}}{2C} T^\text{m} + \frac{D_{,\phi}}{2C}T^\text{m}_{\mu\nu}\nabla^\mu \phi \nabla^\nu \phi - \nabla^\mu \left( \frac{D}{C} T^\text{m}_{\mu \nu} \nabla^\nu \phi \right)\;.
\end{equation}
Here $T^\text{m}$ is the trace of $T^\text{m}_{\mu\nu}$ and a subscript $",\phi"$ stands for derivative with respect to $\phi$.
Finally, the equations of motion for the scalar field are given by the modified Klein-Gordon equation
\begin{equation}
    \square \phi = V_{,\phi} - \Gamma\;.
\end{equation}
We now briefly present the main equations describing the cosmological evolution in such a theory.


\subsection{Background cosmology equations}

Our framework is the usual Friedmann-Lema\^itre-Robertson-Walker (FLRW) metric with spatially flat hypersurfaces:
\begin{equation}
	g_{\mu\nu}dx^\mu dx^\nu = a^2(-d\tau^2 + \delta_{ij}dx^i dx^j)\;.
\end{equation}
As we showed above, the disformal transformation \eqref{disftransf} induces a coupling $\Gamma$ between DM and the scalar field. Considering pressureless DM, so that $T^\text{m} = \rho_\text{m}$ and writing $\Gamma = Q\rho_\text{m}$, with 
\begin{equation}\label{Qstareq}
	Q \equiv -\frac{a^2C_{,\phi} - 2D(3\mathcal H^2\phi' + a^2V_{,\phi} + C_{,\phi}\mathcal H^2\phi'^2/C) + D_{,\phi}\mathcal H^2\phi'^2}{2[a^2C + D(a^2\rho_\text{m} - \mathcal H^2\phi'^2)]} = -\frac{B}{2A}\;,
\end{equation}
where we have defined, for future convenience:
\begin{align}
\label{Adef}	A &\equiv a^2C + D(a^2\rho_\text{m} - \mathcal H^2\phi'^2)\;,\\
\label{Bdef}	B &\equiv a^2C_{,\phi} - 2D(3\mathcal H^2\phi' + a^2V_{,\phi} + C_{,\phi}\mathcal H^2\phi'^2/C) + D_{,\phi}\mathcal H^2\phi'^2\;,
\end{align}
the continuity equation for DM is:
\begin{equation}\label{bgconteq}
	\rho_\text{m}' + 3\rho_\text{m} = -Q\rho_\text{m}\phi'\;,
\end{equation}
whereas the equation of motion for $\phi$ is the following:
\begin{equation}\label{KGequation}
	\phi'' + (2 + \xi)\phi' + \frac{a^2V_{,\phi}}{\mathcal H^2} = \frac{Q\rho_\text{m}a^2}{\mathcal H^2} = 3QM_\text{P}^2\Omega_\text{m}\;,
\end{equation}
where we have introduced the density parameters:
\begin{equation}\label{eq_density_parameters}
	\Omega_\text{m} \equiv \frac{\rho_\text{m}}{\rho_{\rm tot}} = \frac{\rho_\text{m}a^2}{3\mathcal H^2M_\text{P}^2}\;, \qquad \Omega_\phi \equiv \frac{\rho_\phi}{\rho_{\rm tot}} =  \frac{\rho_\phi a^2}{3\mathcal H^2M_\text{P}^2} = \frac{\phi'^2}{6M_\text{P}^2} + \frac{a^2V(\phi)}{3\mathcal H^2M_\text{P}^2}\;.
\end{equation}
In the above equations the prime denotes derivation with respect to the $e$-folds number $N \equiv \ln a$ and we have defined $\xi \equiv \mathcal H'/\mathcal H$, where $\mathcal H \equiv (da/d\tau)/a$ is the conformal Hubble factor.

As usual, the scalar-field density, pressure and equation of state are the following:
\begin{equation}\label{rhophipphi}
	\rho_\phi = \frac{\phi'^2\mathcal H^2}{2a^2} + V(\phi)\;, \qquad p_\phi = \frac{\phi'^2\mathcal H^2}{2a^2} - V(\phi)\;, \qquad w_\phi \equiv \frac{p_\phi}{\rho_\phi}\;,
\end{equation}
 and thus the Friedmann equation can be written as: 
\begin{equation}\label{Friedeq}
	3M_\text{P}^2\mathcal H^2 = (\rho_\text{m} + \rho_\phi)a^2\;, \qquad \Omega_\text{m} + \Omega_\phi = 1\;,
\end{equation}
where we neglect baryons and radiation. 


\subsection{Cosmological perturbations equations}

We focus on scalar perturbations only and write the perturbed FLRW metric in the Newtonian gauge as follows:
\begin{equation}
	g_{\mu\nu}dx^\mu dx^\nu = a^2[-(1 + 2\Psi)d\tau^2 + (1 - 2\Phi)\delta_{ij}dx^i dx^j]\;.
\end{equation}
The spatial traceless part of the Einstein equations tells us that, when no anisotropic stresses are present, the two gravitational potentials are equal: 
\begin{equation}\label{eq:phipsi}
	\Phi = \Psi\;,
\end{equation}
so there is no gravitational slip. The spatial trace of the Einstein field equations is:
\begin{equation}\label{presseq}
	\Phi'' + \left(3 + \xi\right)\Phi' + \left(1 + 2\xi\right)\Phi = \frac{1}{2M_\text{P}^2}\left(\phi'\delta\phi' - \Phi\phi'^2 - \frac{a^2V_{,\phi}\delta\phi}{\mathcal H^2}\right)\;,
\end{equation}
whereas the relativistic Poisson equation is:
\begin{equation}\label{Poissoneq}
	\frac{k^2}{\mathcal H^2}\Phi + 3(\Phi' + \Phi) = -\frac{3}{2}\Omega_\text{m}\delta_\text{m} - \frac{1}{2M_\text{P}^2}\left(\phi'\delta\phi' - \Phi\phi'^2 + \frac{a^2V_{,\phi}\delta\phi}{\mathcal H^2}\right)\;. 
\end{equation}
The continuity and Euler equations for DM are \cite{vandeBruck:2015ida}:
\begin{align}
\label{conteqc}	\delta_\text{m}' &= -\frac{\theta_\text{m}}{\mathcal H} + 3\Phi' + Q\phi'\delta_\text{m} - \phi'\delta Q - Q\delta\phi'\;,\\
\label{euleqc}	\theta_\text{m}' &= -\theta_\text{m} + \frac{k^2\Phi}{\mathcal H} + Q\phi'\theta_\text{m} - \frac{Q}{\mathcal H}k^2\delta\phi\;.
\end{align}
The perturbed Klein-Gordon equation has the following form:
\begin{equation}\label{deltaphieqlna}
	\delta\phi'' + \left(2 + \xi\right)\delta\phi' + \frac{k^2 + a^2V_{,\phi\phi}}{\mathcal H^2}\delta\phi = 4\phi'\Phi' - \frac{2a^2(V_{,\phi} - \rho_\text{m} Q)\Phi}{\mathcal H^2} + \frac{a^2\rho_\text{m}\delta Q}{\mathcal H^2}\;,
\end{equation}
where the perturbed coupling is:
\begin{equation}\label{pertQ}
	\delta Q = -\frac{1}{A}\left(\mathcal B_1\delta_\text{m} + \mathcal B_2\Phi' + \mathcal B_3\Phi + \mathcal B_4\delta\phi' + \mathcal B_5\delta\phi\right)\;, 
\end{equation}
with coefficients \cite{vandeBruck:2015ida}:
\begin{align}
	\mathcal B_1 &\equiv  \frac{B}{2} + a^2DQ\rho_\text{m}\;,\\
\label{B2formula}	\mathcal B_2 &\equiv  3\mathcal H^2D\phi'\;,\\
\label{B3formula}	\mathcal B_3 &\equiv  6\mathcal H^2D\phi' + 2\mathcal H^2D\phi'^2\left(\frac{C_{,\phi}}{C} - \frac{D_{,\phi}}{2D} + Q\right)\;,\\
\label{B4formula}	\mathcal B_4 &\equiv  -3\mathcal H^2D - 2\mathcal H^2D\phi'\left(\frac{C_{,\phi}}{C} - \frac{D_{,\phi}}{2D} + Q\right)\;,\\
\label{B5formula}	\mathcal B_5 &\equiv  \frac{a^2C_{,\phi\phi}}{2} -D(k^2 + a^2V_{,\phi\phi}) - a^2D_{,\phi}V_{,\phi} - 3\mathcal H^2D_{,\phi}\phi'\nonumber\\
	&\qquad- \mathcal H^2D\phi'^2\left[\frac{C_{,\phi\phi}}{C} - \left(\frac{C_{,\phi}}{C}\right)^2 + \frac{C_{,\phi}D_{,\phi}}{CD} - \frac{D_{,\phi\phi}}{2D}\right]\nonumber\\ &\qquad+ (a^2C_{,\phi} + a^2D_{,\phi}\rho_\text{m} - \mathcal H^2D_{,\phi}\phi'^2)Q\;.
\end{align}
In order to find a simple equation ruling the evolution of $\delta_\text{m}$ we present in the next section a set of approximations devised for this scope. The most important one is $k \gg \mathcal H$, which characterises the so-called quasi-static approximation, but also other more specific assumptions shall be needed. 


\section{Approximations}\label{Sec:Approximations}

We find that the following three assumptions are sufficient in order to allow us to find a closed, second-order differential equation for $\delta_\text{m}$: 
\begin{itemize}
	\item \textbf{Approximation 1:} $k/\mathcal H \gg 1$. That is, we consider only sub-horizon scales. Since the sound speed in our case is $c_\text{s}=1$, we also are in the sub-sound-horizon limit.
	
\item \textbf{Approximation 2:} we consider $\Phi'' \sim \Phi' \sim \Phi$ and of the same order of $\Psi \sim \Psi'$, since $\Psi = \Phi$. Moreover we also consider $\delta\phi'' \sim \delta\phi' \sim \delta\phi$. These requirements amount to asking that none of the aforementioned perturbed variables grows too fast, i.e. that there are no instabilities. 

\item \textbf{Approximation 3:} We further ask that $\phi'' \sim \phi'$ and $|Q|M_\text{P} = \mathcal O(1)$. Recall that a quantity is equal to $O(1)$ if it can be at most of order unity. This means that it can also be vanishing.
\end{itemize}
Because of the Friedmann equation constraint \eqref{Friedeq}, $\Omega_\phi \le 1$, then also:
\begin{equation}\label{eq:constraint_on_phip}
	\frac{\phi'^2}{M_\text{P}^2} \le 1\;, \qquad \frac{a^2V}{\mathcal H^2M_\text{P}^2} \le 1\;.
\end{equation}
Note that one of these two quantities (which are positive definite) can be zero, e.g. as in a slow-roll phase for the scalar field, for which $\phi'/M_\text{P} \approx 0$, and their being $\le 1$ means that both are much smaller than $k/\mathcal H$. 

Furthermore, since:
\begin{equation}
	\xi = -\frac{1}{2} - \frac{3}{2}w_\phi\Omega_\phi\;,
\end{equation}
and $|w_\phi| = \mathcal O(1)$, then also $|\xi| = \mathcal O(1)$.

Because of Approximation 2 and 3, from Eq.~\eqref{KGequation} we also have: 
\begin{equation}\label{eq:constraint_on_Vphi}
	\frac{a^2V_{,\phi}}{M_\text{P}\mathcal H^2} = \mathcal O(1)\;,
\end{equation} 
and from Eqs.~\eqref{presseq} and \eqref{Poissoneq} we can also conclude that:
\begin{equation}\label{QSAPoisseq}
	\Phi \sim \delta\phi/M_\text{P}\;, \qquad \frac{k^2}{\mathcal H^2}\Phi = -\frac{3}{2}\Omega_\text{m}\delta_\text{m}\;.
\end{equation}
Finally, we can simplify Eq.~\eqref{deltaphieqlna} as follows:
\begin{equation}\label{deltaphieqlna2}
	\left(k^2 + M^2\right)\delta\phi = a^2\rho_\text{m}Q\delta_\text{m} -\frac{a^2\rho_\text{m}}{A}\left(a^2DQ\rho_\text{m}\delta_\text{m} + \mathcal B_3\Phi + \mathcal B_4\delta\phi' + \mathcal B_5\delta\phi\right)\;,
\end{equation}
where we have used the shorthand notation $M^2 \equiv a^2 V_{,\phi\phi}$. 

In the special case for which $D = 0$, i.e. when we have a pure conformal coupling, the three approximations made are enough in order to obtain an interesting result, as we show in the next section. The case including the disformal part of the transformation is treated in Appendix \ref{Sec:App1}. We do not claim that the above three approximations are all necessary, since the third one could be redundant with the second one. However, we do not go deeper into this mathematical point and simply consider the three approximations as independent.


\section{The pure conformal coupling case}\label{Sec:PureCcase}

Let us from now on set $D = 0$. From Eq.~\eqref{Qstareq} we have then:
\begin{equation}
    Q = -\frac{C_{,\phi}}{2C}\;,
\end{equation}
the terms $\mathcal B_{3,4}$ vanish and we can cast Eq.~\eqref{deltaphieqlna2} as follows:
\begin{equation}\label{deltaphieq3conformalcase}
	\left(k^2 + M^2 - a^2\rho_\text{m}Q_{,\phi}\right)\delta\phi = a^2Q\rho_\text{m}\delta_\text{m}\;.
\end{equation}
Therefore, the following new mass scale appears 
\begin{equation}\label{callM2confcase}
	\mathcal M^2 \equiv - a^2\rho_\text{m}Q_{,\phi} = -3\mathcal H^2M_\text{P}^2\Omega_\text{m} Q_{,\phi} = 3\mathcal H^2M_\text{P}^2\Omega_\text{m}\left(\frac{C_{,\phi\phi}}{2C} + \frac{C_{,\phi}^2}{2C^2}\right)\;,
\end{equation}
depending on the second derivative of $C$ with respect to $\phi$. In the limit $k^2 \gg \mathcal H^2$, this new mass scale is only non-negligible with respect to $k^2$ if $|Q_{,\phi}|M_\text{P}^2$ can be arbitrarily large. At the same time we must take into account that we have assumed $QM_\text{P} = \mathcal O(1)$. Is it possible to find a function $Q(\phi)$ which is limited but whose first derivative is large? The answer is positive: we need to choose a conformal coupling $C$ such that its second derivative is much larger than its first one. This condition is not so restrictive: indeed, it is satisfied in correspondence of extrema of $C$.

Some comments are in order concerning $\mathcal M^2$. First of all, it is not positive definite since $Q_{,\phi}$ can be positive. This means that for some choices of the conformal coupling the new mass scale can even cancel $M^2$. We do not consider this instance in this paper, focusing instead on the quadratic exponential model \eqref{quadraticexponential}, but it could give rise to an interesting phenomenology.

Second, note that if we demand $\mathcal M^2 = 0$ then we obtain from Eq.~\eqref{callM2confcase} that the conformal coupling must be an exponential function of $\phi$ and, since $Q = -C_{,\phi}/(2C)$, then $Q$ has to be a constant. This is an instance that has been extensively considered in the literature, see e.g. \cite{Wetterich:1987fk,Wetterich:1987fm,Wetterich:1994bg,Amendola:1999er, vandeBruck:2015ida}. With such a choice, no new mass scale arises. The above argument works also the other way around, i.e. if $Q$ is constant, then $C$ has to be an exponential function of $\phi$ and thus $\mathcal M^2 = 0$. 
Of course, even if the new mass scale does exist, it might be negligible in the quasi-static approximation, so we need in general another condition on $C(\phi)$ in order for this not to happen. This condition, as we have shown above, is that $\phi$ has to be close to some extrema of the function $C(\phi)$. 

A third point is the following: does the evolution of $\phi$ naturally tend to a minimum of $C(\phi)$? If not, the new mass scale $\mathcal M^2$ is negligible. To see if this is the case, consider Eq.~\eqref{KGequation} and incorporate the term on the right hand side in an effective potential:
\begin{equation}
    \phi'' + (2 + \xi)\phi' + \frac{a^2}{\mathcal H^2}V_{\rm eff,\phi} = 0\;,
\end{equation}
where
\begin{equation}
    V_{\rm eff} \equiv V - \rho_\text{m}\int d\phi Q(\phi) = V + \rho_\text{m}\log \sqrt{C}\;.
\end{equation}
The evolution of $\phi$ tends then towards a minimum of $V_{\rm eff}$, which contains $C$. Evidently, if this minimum of $V_{\rm eff}$, towards which the scalar field rolls down, coincides with a minimum of $C(\phi)$, then the condition by which $\mathcal M^2$ gets very large might be met. 

A fourth and final point is that, as we have shown above, in order to have a large $\mathcal M^2$ we need to be close to the minimum of $C$. So, there is a phase during the cosmological evolution during which this happens and, depending on the effective potential under investigation, this phase could be so short that the emergence of the new mass scale $\mathcal M^2$ could be not easily observable. Here we assume that all the conditions making $\mathcal M^2$ not negligible are met and find in the next subsection the evolution equation for $\delta_\text{m}$ and the role that $\mathcal M^2$ plays in it.

As a concrete example, consider the following coupling:
\begin{equation}\label{quadraticexponential}
	C(\phi) = e^{m_C^{-2}\phi^2}\;.
\end{equation}
It has a minimum for $\phi = 0$ and:
\begin{equation}\label{eq:Q_and_curly_M_in_quadratic_case}
	Q = -\frac{\phi}{m_C^2}\;, \qquad Q_{,\phi} = -\frac{1}{m_C^2}\;, \qquad \mathcal M^2 = 3\Omega_\text{m}\frac{\mathcal H^2M_\text{P}^2}{m_C^2}\;.
\end{equation}
For this model we have $Q_{,\phi} < 0$ and thus a positive new mass scale $\mathcal M^2$. The conditions $|Q|M_\text{P} = \mathcal O(1)$ and $|Q_{,\phi}|M_\text{P}^2 \gg 1$ can be met if:
\begin{equation}\label{conditionphi2model}
    \frac{|\phi| M_\text{P}}{m_C^2} = \mathcal{O}(1)\;, \qquad \frac{M_\text{P}^2}{m_C^2} \gg \frac{1}{3\Omega_\text{m}}\;,
\end{equation}
which are not in contradiction if $\phi$ is close to the minimum $\phi = 0$ (more precisely if $|\phi| \ll M_\text{P}$) and entail:
\begin{equation}
    \mathcal{M}^2 \gg \mathcal H^2\;,
\end{equation}
since $\Omega_\text{m} \le 1$. Therefore, if the above conditions are met, the new mass scale $\mathcal M^2$ is not negligible in the quasi-static approximation.

For the particular case of the quadratic exponential coupling of Eq.~\eqref{quadraticexponential}, if the scalar field potential is simply taken to be $a^2V = M^2\phi^2/2$, then the effective potential amounts to $a^2V_{\rm eff} = \frac{1}{2}(M^2 + \mathcal{M}^2) \phi^2$ and indeed we expect an evolution such that $\phi\to 0$. Notice that this particular model will not drive acceleration. We employ it only to provide a concrete example that realises a weaker gravity.

\subsection{The evolution equation for the DM density contrast}

Thanks to Eq.~\eqref{deltaphieq3conformalcase}, which we cast as follows:
\begin{eqnarray}\label{KGeqQSAter}
	\delta\phi = \frac{a^2Q\rho_\text{m}}{k^2 + M^2 + \mathcal M^2}\delta_\text{m}\;,
\end{eqnarray}
and thanks to the approximations made, we are now in the position to obtain a second-order differential equation for $\delta_\text{m}$, which we cast in the following form:
\begin{equation}
	\delta_\text{m}'' + F\delta_\text{m}' - \frac{3}{2}\Omega_\text{m}(1 + Y)\delta_\text{m} =  0\;,
\end{equation}
with $F$ being a friction term and $Y$ the effective gravitational coupling. Within our approximation scheme, the perturbed continuity equation \eqref{conteqc} is simplified as follows:
\begin{equation}\label{conteqcQSA}	
    \delta_\text{m}' = -\frac{\theta_\text{m}}{\mathcal H} - Q'\delta\phi\;.
\end{equation}
Here we cannot neglect the last term since $Q_{,\phi}$ is large, which is why $Q'$ might in principle be large as well. This is a crucial consideration, that leads to our novel result. For example, in the quadratic-exponential model of Eq.~\eqref{quadraticexponential} we have:
\begin{equation}
   |Q'|M_\text{P} = \frac{|\phi'|M_\text{P}}{m_C^2}\;.\label{eq:apprphip}
\end{equation}
This can be large without being in contradiction with the previous two conditions of Eq.~\eqref{conditionphi2model} because $|\phi'|/M_\text{P} = \mathcal O(1)$. In other words, for the quadratic exponential model we need $\phi$ to be close to zero but not its derivative.

Therefore, keeping the term $Q'\delta\phi$, using Eq,~\eqref{KGeqQSAter} and the Euler equation, we obtain:
\begin{align}
\label{eq:QSAdeltam} \delta_\text{m}' &= -\frac{\theta_\text{m}}{\mathcal H} + \frac{\mathcal M^2}{k^2 + M^2 + \mathcal M^2}Q \phi'\delta_\text{m}\;,\\
\label{eq:QSAthetam} \frac{\theta_\text{m}'}{\mathcal H} &= -\frac{\theta_\text{m}}{\mathcal H} - \frac{3}{2}\Omega_\text{m}\delta_\text{m} + Q\phi'\frac{\theta_\text{m}}{\mathcal H} - 3M_\text{P}^2Q^2\Omega_\text{m}\frac{k^2}{k^2 + M^2 + \mathcal M^2}\delta_\text{m}\;,
\end{align}
and combining them we arrive at the following result:
\begin{align}
	\delta_\text{m}'' &+ \left(1 + \xi - Q\phi' - Q\phi'\frac{\mathcal M^2}{k^2 + M^2 + \mathcal M^2}\right)\delta_\text{m}' \\ 
	&-\frac{3}{2}\Omega_\text{m}\left[1 + \frac{2M_\text{P}^2Q^2k^2 + \frac{(1 + \xi - Q\phi')Q\phi' + (Q\phi')'}{3\Omega_\text{m}/2}\mathcal M^2}{k^2 + M^2 + \mathcal M^2} + \left(\frac{\mathcal M^2}{k^2 + M^2 + \mathcal M^2}\right)'\frac{Q\phi'}{3\Omega_\text{m}/2}\right]\delta_\text{m} =  0\;\nonumber.
\end{align}
Hence, we can read off:
\begin{align}
	F &= 1 + \xi - Q\phi' - Q\phi'\frac{\mathcal M^2}{k^2 + M^2 + \mathcal M^2}\;, \label{eq_expression_for_F}\\
	Y &= 2M_\text{P}^2Q^2\frac{k^2 + g\mathcal M^2}{k^2 + M^2 + \mathcal M^2} +\left(\frac{\mathcal M^2}{k^2 + M^2 + \mathcal M^2}\right)'\frac{2Q\phi'}{3\Omega_\text{m}}\;,
\end{align}
where
\begin{equation}
	g \equiv \frac{(1 + \xi - Q\phi')Q\phi' + (Q\phi')'}{3\Omega_\text{m}M_\text{P}^2Q^2}\;.
\end{equation}
Note that, as anticipated, the friction term now gains a scale-dependence, due to the new mass scale $\mathcal M^2$. 

We can rewrite the above $Y$ as follows:
\begin{equation}\label{eq_expression_Y}
	Y = 2M_\text{P}^2Q^2\frac{k^4 + \alpha_2k^2 + \alpha_0}{(k^2 + M^2 + \mathcal M^2)^2} = 2M_\text{P}^2Q^2\frac{k^4 + \alpha_2k^2 + \alpha_0}{k^4 + \beta_2k^2 + \beta_0}\;,
\end{equation}
with the following definitions:
\begin{eqnarray}
\label{alpha2}	\alpha_2 &\equiv& \frac{\phi'}{3\Omega_\text{m} M_\text{P}^2Q}(\mathcal M^2)' + \bar M^2 + g\mathcal M^2\;,\label{eq_definition_alpha2}\\
\label{alpha0}	\alpha_0 &\equiv& \frac{\phi'}{3\Omega_\text{m} M_\text{P}^2Q}\left[(\mathcal M^2)'M^2 - (M^2)'\mathcal M^2\right] + \mathcal M^2\bar M^2g\;,\label{eq_definition_alpha0}\\
\label{beta}   \beta_2 &\equiv& 2\bar M^2\;, \qquad \beta_0 \equiv \bar M^4\;, \label{eq_definition_beta2}
\end{eqnarray}
where we have introduced the combined mass scale:
\begin{equation}
    \bar M^2 \equiv M^2 + \mathcal M^2\;.
\end{equation}
Note that in the slow-roll limit in which $\phi'$ and $\phi''$ are small, one recovers the standard case of Eq.\,\,(\ref{eq:qsa1}),
\begin{equation}
    Y=2M_\text{P}^2 Q^2 \frac{k^2}{k^2+M^2}\;.
\end{equation}
Furthermore, as we have mentioned above, the case $\mathcal M = 0$ corresponds to a constant $Q$ so that one recovers the standard case as well.

\subsection{Weak gravity}

The new $Y$ in Eq. \eqref{eq_expression_Y} seems to have a novel behaviour, characterised by a ratio of fourth-order polynomial in $k$. On the other hand, we must recall that the function $g$ contains a term proportional to $Q'$. We are interested in a specific limit which is characterised by a weakening of gravity. In order to obtain this limit, we assume  $\phi' \approx M_\text{P}$ and therefore $Q'$ to be large and dominating all the other terms in $g$ itself. In other words we assume that the quantity 
\begin{equation}
    g \simeq \frac{Q'\phi'}{3\Omega_\text{m}M_\text{P}^2Q^2}\;,
\end{equation}
is large with respect to unity. Since 
\begin{equation}
    (\mathcal M^2)' = 2\xi\mathcal M^2 + \frac{\Omega_\text{m}'}{\Omega_\text{m}}\mathcal M^2 + \frac{Q_{,\phi\phi}\phi'}{Q_{,\phi}}\mathcal M^2\;,
\end{equation}
unless some special cases are considered, we have in general that $(\mathcal M^2)'$ is of the same order of $\mathcal M^2$. Therefore, the coefficients of Eq. (\ref{eq_expression_Y})  can be simplified as follows:
\begin{eqnarray}
\alpha_2 &\simeq& M^2 + g\mathcal M^2\;,\\
	\alpha_0 &\simeq&  g\mathcal M^2\bar M^2\;,
\end{eqnarray}
where we have also assumed that $(M^2)'$ is of the same order of $M^2$. Putting them into $Y$, one gets:
\begin{equation}
    Y \simeq 2M_\text{P}^2Q^2\frac{k^4 + (M^2 + g\mathcal M^2)k^2 + g\mathcal M^2\bar M^2}{(k^2 + \bar M^2)^2}\;.
\end{equation}
Since for $\phi' \approx M_\text{P}$ the contribution $g\mathcal M^2k^2$ is much larger than $k^4 + M^2k^2$,  this simplifies to:
\begin{equation} \label{ywg}
    Y \simeq 2M_\text{P}^2Q^2g\mathcal M^2\frac{1}{k^2 + \bar M^2} = -2M_\text{P}^2(Q')^2\frac{1}{k^2 + \bar M^2}\;.
\end{equation}
In the quadratic exponential case this corresponds to:
\begin{equation}
    Y \simeq -2M_\text{P}^2\frac{\mathcal H^2(\phi')^2}{m_C^2}\frac{1}{k^2 + \bar M^2} = -2M_\text{P}^2Q^2\frac{\mathcal H^2(\phi')^2}{\phi^2}\frac{1}{k^2 + \bar M^2}\;.
\end{equation}
In real space, the potential is
\begin{equation}
    V(r)=-\frac{G_\text{N} m}{r}\left[1-\frac{2M_\text{P}^2(Q')^2}{\bar M^2} (1-e^{-\bar M r})\right]\;,
\end{equation}
which  generates weaker gravity (perhaps even repulsive) on large scales.

Summarising, we find a new behaviour of the effective gravitational coupling $Y$, characterised by a Yukawa-like contribution counteracting the Newtonian potential on large scales when the conditions $|Q_{,\phi}|M_\text{P}^2 \gg 1$ and $|Q'|M_\text{P} \gg 1$ are met and the approximations made in Sec.~\ref{Sec:Approximations} are valid. In Appendix \ref{Sec:App3} we show by a numerical solution of the full set of background and perturbation equations that the QSA is valid and that the transient weaker gravity phenomenon  indeed occurs when the above conditions are fulfilled.

An interesting question is whether such novel behaviour could be found also in the Jordan frame, where e.g. Horndeski theory is formulated. In Appendix \ref{Sec:App2} we present the transformations between the two frames concerning the model under investigation and comment on how the approximation made in the Einstein frame should be modified in the Jordan one. We do not perform an explicit calculation in the Jordan frame, leaving it as a future task. 


\section{Conclusions}\label{Sec:Conclusions}

In this paper we have studied a simple version of a scalar-tensor theory of gravity for which dark matter is minimally coupled to a conformally transformed geometry. We have shown that for a non-constant coupling function $Q$ a new mass scale arises. If such conformal coupling presents a minimum and the dynamics of the scalar field tend to that minimum, then the growth of small dark matter fluctuations is characterised by an effective gravitational coupling which is smaller than that in the standard case. At the same time, the friction term in the fluctuation equation acquires a novel $k$-dependent term. We have identified the conditions under which such a novel behaviour takes place and proposed a model for the conformal coupling function, a quadratic exponential function of the scalar field, which has the potentiality to display it. In App. \ref{Sec:App2} we confirm numerically some of the features of our model,  leaving a more systematic investigation as a future project. It is also interesting to check whether the novel behaviour shows itself in the Jordan frame. In Appendix \ref{Sec:App3} we present the transformations between the two frames and suggest that the quasi-static approximation that we have considered here must be properly translated into the Jordan frame in order to have a similar behaviour for $Y$. We also leave as a future development a deeper investigation of this point. 


\begin{acknowledgments}
This study was financed in part by the \emph{Coordena\c{c}\~ao de Aperfei\c{c}oamento de Pessoal de N\'ivel Superior} - Brazil (CAPES) - Finance Code 001. MW thanks the DFG for support through the Research Training Group “Particle Physics beyond the Standard Model” (GRK 1940). OFP thanks the Alexander von Humboldt foundation for funding and the Institute for Theoretical Physics of Heidelberg University for kind hospitality.  LA thanks Shinji Tsujikawa for interesting discussions on this topic.
\end{acknowledgments}


\appendix

\section{Including a disformal coupling}\label{Sec:App1}

In this Appendix we consider also the disformal term. Keeping $D \neq 0$, let us see under which conditions we can neglect the contributions $\mathcal B_2\Phi'$, $\mathcal B_3\Phi$ and $\mathcal B_4\delta\phi'$ with respect to $a^2DQ\rho_\text{m}\delta_\text{m}$ in Eq.~\eqref{deltaphieqlna2}. Moreover, let us investigate how we can simplify the $\mathcal B_5\delta\phi$ contribution under the same conditions.

Based on our approximations of Sec.~\ref{Sec:Approximations}, we can see that:
\begin{equation}
	\mathcal B_2\Phi',\; \mathcal B_3\Phi,\; \mathcal B_4\delta\phi' \sim \mathcal H^2DM_\text{P}\;,
\end{equation}
because
\begin{equation}
	\left|\frac{C_{,\phi}}{C}\right|M_\text{P} = \left|\frac{C'}{C\phi'}\right|M_\text{P} = \mathcal O(1)\;,
\end{equation}
and similarly for the logarithmic derivative of $D$. So, because of Approximation 1 ($k^2 \gg \mathcal H^2$) these terms are negligible with respect to the $D(k^2 + M^2)\delta\phi$ one.

Within $\mathcal B_5$ we have the following approximations:
\begin{align}
	\frac{a^2C_{,\phi\phi}}{2} \sim \frac{C}{M_\text{P}^2}\;, \qquad a^2D_{,\phi}V_{,\phi},\; 3\mathcal H^2D_{,\phi}\phi' \sim \mathcal H^2D\;, \nonumber\\
	\mathcal H^2D\phi'^2\left[\frac{C_{,\phi\phi}}{C} - \left(\frac{C_{,\phi}}{C}\right)^2 + \frac{C_{,\phi}D_{,\phi}}{CD} - \frac{D_{,\phi\phi}}{2D}\right] \sim \mathcal H^2D\;, \quad a^2C_{,\phi}Q \sim \frac{C}{M_\text{P}^2}\;, \nonumber\\ a^2D_{,\phi}\rho_\text{m}Q, \mathcal H^2D_{,\phi}\phi'^2Q \sim \mathcal H^2D\;.
\end{align}
All the terms similar to $\mathcal H^2D$ can be neglected with respect to $Dk^2$ and we are left just with:
\begin{equation}\label{B5gencase}
	\mathcal B_5 = \frac{a^2C_{,\phi\phi}}{2} - D(k^2 + M^2) + a^2C_{,\phi}Q\;,
\end{equation}
with the two terms containing $C$ (which, by the way, are the only ones that survive if we take $D = 0$, see Sec.~\ref{Sec:PureCcase}) providing a new scale $k_{CD}^2$ for which:
\begin{equation}
	k_{CD}^2 \equiv \frac{a^2C_{,\phi\phi}}{2D} + \frac{a^2C_{,\phi}Q}{D} \sim \frac{C}{DM_\text{P}^2}\;.
\end{equation}
If $k_{CD} \sim \mathcal H$ then we can take $\mathcal B_5 = -D(k^2 + M^2)$ because of our Approximation 1 and in this case we obtain the standard coupling $Y = 2M_\text{P}^2Q^2k^2/(k^2 + M^2)$. On the other hand, if $k_{CD} \gg \mathcal H$, we can retain the conformal terms and combining Eq.~\eqref{deltaphieqlna2} and Eq.~\eqref{B5gencase} one obtains:
\begin{equation}\label{deltaphieq3}
	\left(k^2 + M^2\right)\delta\phi + \frac{a^2\rho_\text{m}(a^2C_{,\phi\phi}/2
	+ a^2C_{,\phi}Q)}{A - a^2\rho_\text{m}D}\delta\phi = a^2Q\rho_\text{m}\delta_\text{m}\;.
\end{equation}
We can thus introduce again a mass scale, similar to the one defined in Eq.~\eqref{callM2confcase}, but now containing the disformal contribution:
\begin{equation}\label{callM2}
	\mathcal M^2 \equiv \frac{a^2\rho_\text{m}(a^2C_{,\phi\phi}/2
	+ a^2C_{,\phi}Q)}{A - a^2\rho_\text{m}D} = \frac{a^2\rho_\text{m}(a^2C_{,\phi\phi}/2
	+ a^2C_{,\phi}Q)}{a^2C - D\mathcal H^2\phi'^2}\;.
\end{equation}
However, since $k_{CD} \gg \mathcal H$, then the above denominators can be simplified as
\begin{equation}
    a^2C - D\mathcal H^2\phi'^2 \approx a^2C\;,
\end{equation}
because $|C| \gg |D|M_\text{P}^2\mathcal H^2$ and $\phi'^2/M_\text{P}^2 \sim \Omega_\phi \sim \mathcal O(1)$. Therefore, the mass scale $\mathcal{M}^2$ has the same expression as in Eq.~\eqref{callM2confcase}, for the conformal case. Moreover, if $|C| \gg |D|M_\text{P}^2\mathcal H^2$, then $Q$ from Eq.~\eqref{Qstareq} can be simplified as $Q = -C_{,\phi}/(2C)$, i.e. the same as we have in the purely conformal case.

Hence, we can conclude that a new mass scale $\mathcal{M}^2$ appears only when the disformal contribution is subdominant with respect to the conformal one and $C(\phi)$ is not an exponential function, entailing then a non-constant $Q$.  


\section{Numerical analysis}\label{Sec:App2}

The quasi-static approximation introduced earlier defines a regime in which, when the additional assumptions on weak gravity are met, the growth of matter perturbations is hindered by the negative value of the effective gravitational potential $Y$ in Eq. (\ref{ywg}). Due to the oscillations of the dynamical scalar field, we expect this behaviour to exist only as a transient phenomenon. To further investigate this point, we proceed with a numerical analysis of the conformal coupling model, setting the conformal coupling function to be a quadratic exponential of $\phi$ (as in Eq. (\ref{csqexp})) and the scalar field potential to be quadratic in $\phi$:
\begin{gather}
    C(\phi)=C_0e^{m_c^{-2} \phi^2}\;,\\
    V(\phi)=\frac{1}{2}m_v^2\phi^2\;,
\end{gather}
with $C_0$ set to unity. The model is then entirely specified in terms of the sole two parameters $m_c$ and $m_v$ and the coupling function takes the form $Q(\phi)=-m_c^{-2}\phi$. All the variables and parameters discussed in this section are taken to be dimensionless quantities. The conformal Hubble function $\mathcal{H}$ is normalized to 1 at the present cosmological time. The comoving wave number $k$ is also defined to be dimensionless by dividing it by $H_0$.

\begin{figure}[t]
    \includegraphics[scale=0.5]{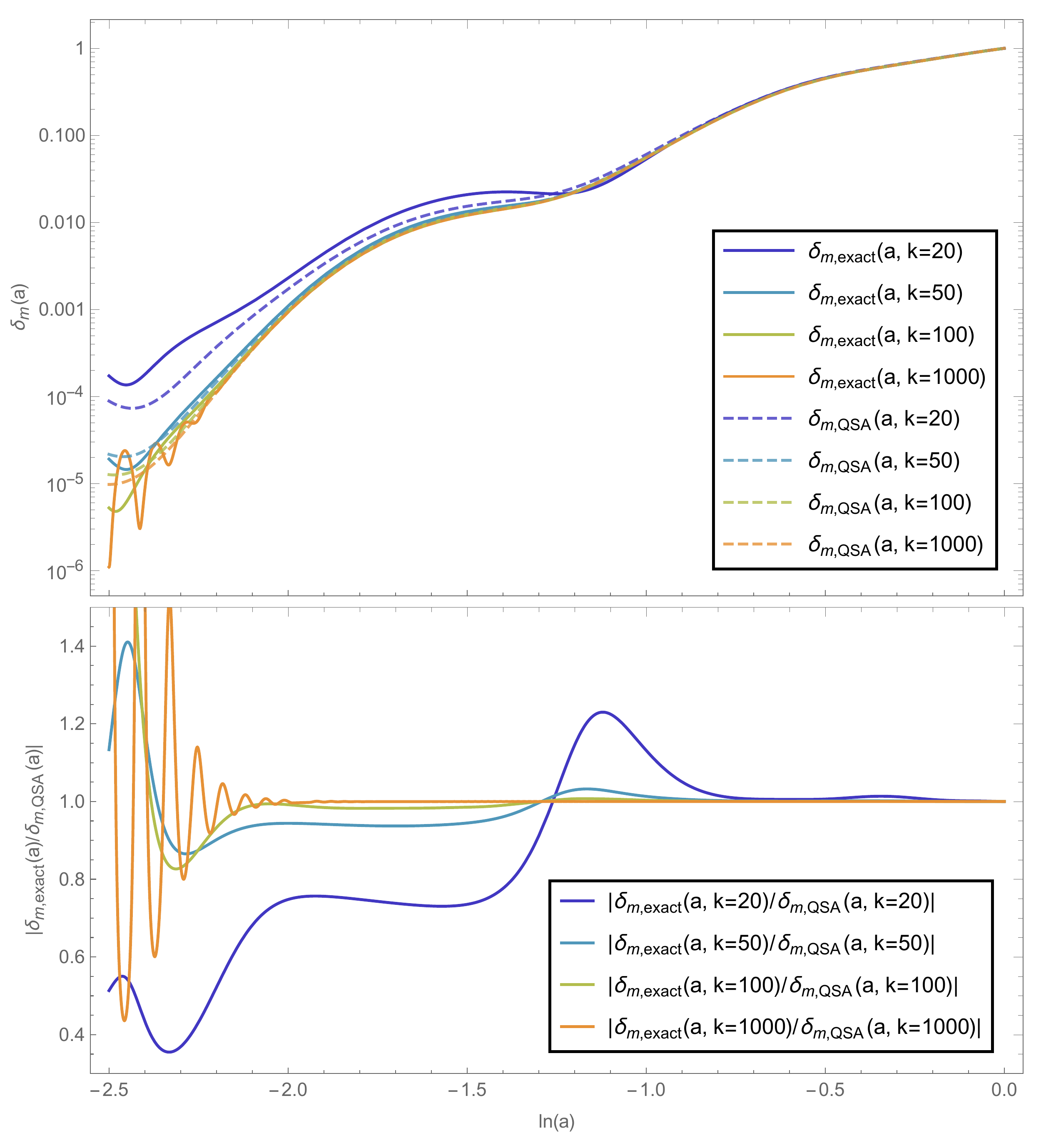}
    \caption{Matter-density contrast in the $m_c^{-2} = 4$ case for $k= 20, \; 50, \; 100, \;1000$. The comparison is between the exact solutions (full lines, labelled as ``exact'') and the quasi-static approximated solutions (dashed lines, labelled as ``QSA''). Every perturbation contrast $\delta_\text{m}(a)$ has been normalised to its value at the present cosmological time. The second panel shows the ratio $|\delta_\text{m,exact}(a)/\delta_\text{m,QSA}(a)|$ for the chosen values of $k$.}
    \label{fig.pertk}
\end{figure}

As a first step, we perform the numerical computation of the background evolution of $\rho_\text{m}(a)$ and $\phi(a)$ by solving the system of differential equations in Eqs. \eqref{bgconteq} and \eqref{KGequation}. Boundary conditions are applied at both ends of the time interval, namely by fixing the initial values $\phi(a_\text{in})=2$, $\phi'(a_\text{in})=0$ and the final (i.e. present-day) values $\Omega_\text{m}(a_{0})=0.31$ and $\Omega_{\phi}(a_0)=0.69$. The system is then solvable via a shooting method. The parameter $m_v$ is left unspecified and its value is obtained automatically by the shooting method as a fitting parameter. The evolution encompasses the time-span $\ln{(a)} \in [-4, 0]$. This setup allows to have $Q=\mathcal{O}(1)$ throughout the cosmic history if we set $m_c^{-2}=\mathcal{O}(1)$ while $\phi(a)=\mathcal{O}(1)$. Table \ref{tab.inconbg} contains the values of the parameters $m_c$ and $m_v$ used in the analysis. From the Klein-Gordon equation for $\phi$ (Eq. (\ref{KGequation})) we see that the oscillation frequency depends on the sum of the two terms $M^2=m_v^2a^2$  and $\mathcal{M}^2=m_c^{-2}a^2\rho_\text{m}$. Hence, we expect higher frequencies for higher values of $m_v$. 
\begin{table}[t]
    \centering
    \begin{tabular}{>{\centering\arraybackslash}p{1cm} | >{\centering\arraybackslash}p{2cm} >{\centering\arraybackslash}p{2cm}}
        Model & $m_c^{-2}$    & $m_v$   \\
         \hline
        1        & $4$      & $10.6778$  \\
        2        & $3$      & $27.5699$  \\
        3        & $2.5$    & $43.7896$  \\
        4        & $2$      & $58.7792$  \\
    \end{tabular}
    \caption{Pairs of parameters $m_c$ and $m_v$ used in the present analysis.}
    \label{tab.inconbg}
\end{table}
The second step consists in computing the evolution of perturbations. We integrate the full system in Eqs. \eqref{Poissoneq}, \eqref{conteqc}, \eqref{euleqc} and \eqref{deltaphieqlna} with the following choice of initial conditions for all the cases: $\delta_\text{m}(a_\text{in})=1$, $\delta \phi(a_\text{in})=1$, ${\delta \phi}'(a_\text{in})=1$, $\theta_\text{m}(a_\text{in})=1$, where we set $a_\text{in}=e^{-2.5}$ and the initial value $\Phi(a_\text{in})$ is obtained from the constraint Eq. (\ref{Poissoneq}). The quasi-static approximated system in Eqs. \eqref{eq:QSAdeltam} and \eqref{eq:QSAthetam} is solved analogously. We now focus on the comparison between the density contrast $\delta_\text{m}$ in the exact solutions and in the quasi-static ones. Fig. (\ref{fig.pertk}) inspects the quality of the quasi-static approximation: for a single value of the conformal-coupling parameter $m_c^{-2} = 4$ it shows different matter-density contrasts for different values of $k$. The condition $k/\mathcal{H}>1$ is met for each solution but, as one would expect, the overall convergence of the quasi-static solution to the exact one is better for larger values of $k$. This is particularly evident at early times, since $\mathcal{H}(a)$ is characterised by a monotonically decreasing behaviour. We can guarantee the early-time matching between the two kinds of solutions only for $k \geq 100$. Similar considerations hold for the other values of the coupling parameter.

\newpage

\begin{figure}[t]
    \includegraphics[scale=0.5]{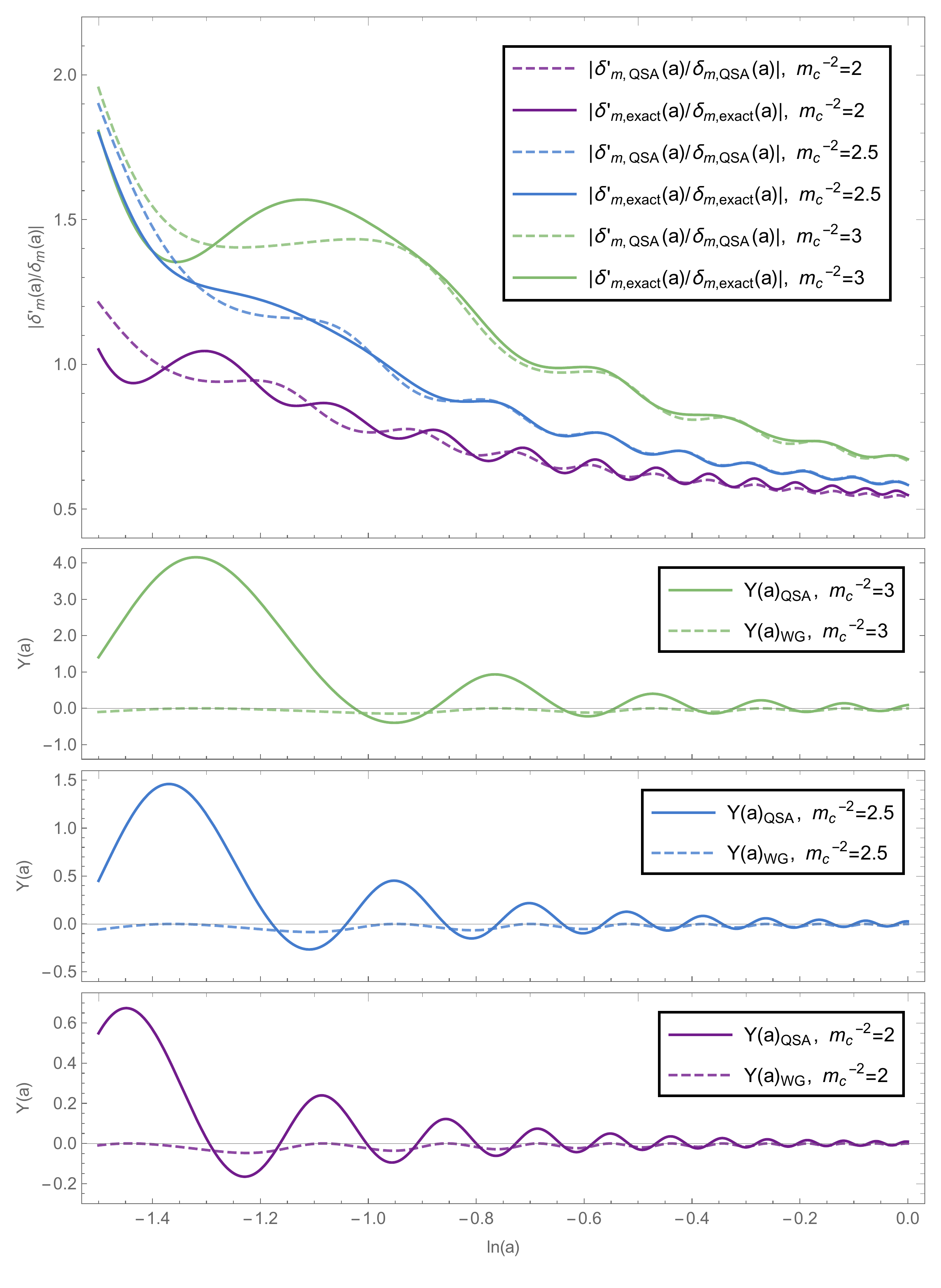}
    \caption{The first panel shows the behaviour of the absolute value of the ratio $\delta_\text{m}'(a)/\delta_\text{m}(a)$ of the matter density contrast $\delta_\text{m}(a)$. The comparison is carried out for $m_c^{-2}=2,\;2.5,\;3$ at fixed $k=20$ in the interval $\ln{(a)} \in [-1.5, 0]$. Here ``QSA'' indicates the quasi-static approximation (dashed lines) while the full lines are the exact numerical solutions.  The second, third and fourth panel refer individually to the cases $m_c^{-2}=2,\;2.5,\;3$ and show simultaneously the behaviour of $Y$ in the quasi-static approximation and in the weak-gravity regime.}
    \label{fig.weakgrav}
\end{figure}

It will then be important to focus on the interval $\ln{(a)} \in [-1.5, 0]$ and choose $k \geq 20$ in order to test the novel behaviour of $Y$.
In Fig. (\ref{fig.weakgrav}) we consider three different coupled models for $k=20$.  

\newpage

From the first panel we can observe two features of the perturbation growth: a local onset of the weak-gravity regime and a long-term suppression of the perturbation growth. The oscillations of the coupling function are in phase with the scalar field oscillations due to the pure conformal coupling $Q(a)=-m_c^{-2}\phi(a)$. These are clearly observable when inspecting the ratio $|\delta_\text{m}'(a)/\delta_\text{m}(a)|$. In particular, when $\phi$ approaches the minimum of the  potential $V$, its first derivative becomes much larger than the value of the scalar field, thus triggering a weak-gravity phase. This phenomenon appears as a series of shorter and shorter intervals during which the growth of perturbations is suppressed as predicted in Eq. (\ref{ywg}). In the long term, we notice that for the cases $m_c^{-2}=2, \; 2.5$ the overall growth is hindered by the periods of weak-gravity.  The advent of the weak-gravity regime can also be inferred from the third panel of Fig. (\ref{fig.weakgrav}): when the overall factor $Q^2$ in the expression for $Y$ reaches its minima (corresponding to the minima of $\phi$), the weak-gravity conditions are fulfilled and the approximated expression of $Y$ in Eq. (\ref{ywg}) matches the quasi-static expression in Eq. (\ref{eq_expression_Y}).


\section{Transformation to the Jordan frame}\label{Sec:App3}

We discuss here the transformation rules between Einstein and Jordan frame, and whether the newly found behaviour in the gravitational coupling could be obtained. Let us consider the pure conformal case only. We use notation and approach as in Ref.~\cite{vandeBruck:2015ida}. The quantities with tilde are in the Jordan frame :
\begin{equation}
    T^{\mu\nu} = C^3T^{\mu\nu}\;, \quad d\tilde s^2 = Ca^2(\tau)(-d\tau^2 + \delta_{ij}dx^idx^j) = Cds\;.
\end{equation}
Therefore,
\begin{equation}
    \tilde u^\mu = \frac{1}{\sqrt{C}}u^\mu\;, \quad \tilde H = \frac{1}{\sqrt{C}}\left[H + \frac{C_\phi}{2C}\frac{\dot\phi}{a}\right]\;, \quad \tilde\rho = \frac{\rho}{C^2}\;, \quad \tilde P = \frac{P}{C^2}\;, \quad \tilde w = w\;.
\end{equation}
In particular, for the Hubble factor we have that, considering that $\tilde a = a\sqrt{C}$:
\begin{equation}
    \frac{\tilde{\mathcal{H}}}{\tilde a} = \frac{1}{a\sqrt{C}}\left[\mathcal H + \frac{C_\phi}{2C}\phi'\mathcal H\right]\;, \quad\longrightarrow\quad  \tilde{\mathcal{H}} = \mathcal H\left[1 - Q\phi'\right]\;.
\end{equation}
Therefore, the condition $k^2 \gg \mathcal H^2$ in the Einstein frame corresponds to \begin{equation}
    \tilde k^2 \gg \mathcal H^2 = \frac{1}{(1 - Q\phi')^2}\tilde{\mathcal{H}}^2\;,
\end{equation}
and moreover
\begin{equation}
    \tilde{\mathcal{H}}' = \mathcal{H}'\left[1 - Q\phi'\right] - \mathcal{H}(Q\phi')'\;.
\end{equation}
Here we see an important difference between the two frames. In the Einstein frame, the term $|Q'|M_\text{P}$ has to be very large in order to make the new mass scale $\mathcal M^2$ not negligible. But then this means that $\tilde{\mathcal{H}}'$ in the Jordan frame has to be very large as well, which is an uncommon requirement for the quasi-static approximation.

This simple argument suggests that the novel scale-dependence that we found here in the Einstein frame can be obtained in the Jordan one only at the price of some extra requirements which might not seem natural. 


\bibliographystyle{unsrturl}
\bibliography{Bibliography}

\end{document}